# On "Exact analytical calculations of thermodynamic functions of gaseous substances, R. Khordad, A. Avazpour, and A. Ghanbari, *Chemical Physics* 517 (2019) 30".

I.H. Umirzakov

Institute of Thermophysics, Lavrentev prospect, 1, 630090 Novosibirsk, Russia

e-mail: cluster125@gmail.com

**Abstract**

It is shown that that the exact solution of the Schrodinger equation for the molecule consisting of two atoms interacting via improved Tietz potential, the analytical expressions for the energy of the rotational-vibrational levels, partition function, free energy, mean energy and specific heat, obtained from the solution in the paper "Exact analytical calculations of thermodynamic functions of gaseous substances", R. Khordad, A. Avazpour, and A. Ghanbari, *Chemical Physics* 517 (2019) 30, are incorrect.

**Introduction**

The improved Tietz interaction potential for the diatomic molecules is given by [1]

$$V(r) = D_e\left(1 - \frac{e^{\alpha r_e} + q}{e^{\alpha r} + q}\right)^2 - D_e, \tag{1}$$

where $D_e$ is the dissociation energy, $D_e > 0$, $r$ is the interatomic separation, $r \geq 0$, $r_e$ is the equilibrium bond length, $r_e > 0$, $q$ and $\alpha$ are the adjustable parameters, $\alpha$ governs the range of interaction, $\alpha > 0$, and $e^{\alpha r_e} + q \neq 0$. Here the energy is measured from the minimum of the potential well, so $V(r_e) = -D_e$ and $V(r \to \infty) = 0$.

The analytical expression for the energy of the rotational-vibrational levels of the diatomic molecule for the improved Tietz potential was obtained in [1] from the exact solution of the stationary Schrodinger equation. The expression was used to obtain analytical expressions for the partition function, free and mean energies and specific heat of diatomic molecules, and the thermodynamic properties of the diatomic molecules $HCl$, $HF$, $DF$, $BBr$, $CO$ and $NO$ calculated from above expressions were compared with the experimental data.

We prove in the present paper that that the exact solution of the Schrodinger equation for the molecules consisting of two atoms interacting via improved Tietz potential, the analytical expressions for the energy of the rotational-vibrational levels, partition function, free energy, mean energy and specific heat obtained in [1] are incorrect.

It is necessary to give three comments before the proof:

1. The detailed analysis of the derivation of Eq. 15 [1] shows that the term $D_e(e^{2\alpha r_e} - q^2)^2$ in the square brackets in the numerator at the right hand side of Eq. 15 [1] must be replaced by $D_e(e^{2\alpha r_e} - q^2)$.

2. The values of the dissociation energy of the diatomic molecules $HCl$, $DF$, $BBr$ and $CO$ given in Table 1 [1] are incorrect. Therefore the thermodynamic functions of these molecules which were presented in Tables 2-4 and Figures 1-3 [1] can be incorrect.

3. It is easy to see that Eq. 8 [1] (the Schrodinger equation) is incorrect, and it is necessary replace the incorrect terms $e^{\alpha r_e} + q$ and $(e^{\alpha r_e} + q)^2$ by the correct ones $e^{\alpha r} + q$ and $(e^{\alpha r} + q)^2$, respectively, in the denominators of the third and fourth terms in the square brackets at the right hand side of Eq. 8 [1].

**The proof**

According to [1] the energy for zero rotational quantum number $J = 0$ is given by

$$E_{v, J=0} = -\frac{\hbar^2 \alpha^2}{2\mu}(A - B)^2, \qquad (2)$$

$$A = \frac{2\mu D_e}{\hbar^2 \alpha^2 q^2} \cdot \frac{e^{2\alpha r_e} - q^2}{2v + 1 - \sqrt{1 + \frac{8\mu D_e}{\hbar^2 \alpha^2 q^2}(e^{\alpha r_e} + q)^2}}, \qquad (3)$$

$$B = \frac{1}{4}\left(2v + 1 - \sqrt{1 + \frac{8\mu D_e}{\hbar^2 \alpha^2 q^2}(e^{\alpha r_e} + q)^2}\right), \qquad (4)$$

where $\hbar$ is the Planck's constant, $\mu$ is the reduced mass, $v$ is the vibrational quantum number and $q \neq 0$.

The energy of the quantum levels must be greater than the minimal value of the potential energy [2] so $E_{v,0}$ must obey the condition

$$E_{v,0} > -D_e. \qquad (5)$$

Therefore if the exact solution of the Schrodinger equation is correct than the solution must obey the condition 5, and if the solution does not obey the condition 5 than the solution is incorrect.

We will consider continuous potential at $r \geq 0$. As evident from Eq. 1 the continuity of the potential requires $q \geq -1$. So the parameter $q$ must obey the inequalities $e^{\alpha r_e} + q \neq 0$, $q \geq -1$ and $q \neq 0$.

According to [1] the analytical expression given by Eq. 2 for the energy of the rotational-vibrational levels of the diatomic molecule for the improved Tietz potential is obtained from the exact solution of the stationary Schrodinger equation, and the solution and Eq. 2 are valid for values of the parameter $q$ obeying the conditions $e^{\alpha r_e} + q \neq 0$, and $q \neq 0$.

If it is proved that the Eq. 2 does not obey the inequality 5 for all above values of the parameter $q$ then Eq. 2 and, hence, the solution of the Schrodinger equation, are incorrect.

We obtain from Eqs. 2-4 and inequality 5

$$\left(\frac{\lambda(a^2-1)}{2v+1-\sqrt{1+\lambda(a+1)^2}} - 2v - 1 + \sqrt{1+\lambda(a+1)^2}\right)^2 < 4\lambda, \tag{6}$$

where $\lambda = 8\mu D_e/\hbar^2\alpha^2$, $\lambda > 0$ and $a = e^{\alpha r_e}/q$. Using the conditions $q \geq -1$, $q \neq 0$ and $e^{\alpha r_e} + q \neq 0$ we obtain from $a = e^{\alpha r_e}/q$ that $a \leq -e^{\alpha r_e}$ and $a > 0$.

Let us consider the case of zero vibrational number: $v = 0$. We have from the inequality 6

$$\left(\frac{\lambda(a^2-1)}{1-\sqrt{1+\lambda(a+1)^2}} - 1 + \sqrt{1+\lambda(a+1)^2}\right)^2 < 4\lambda, \tag{7}$$

We have from the inequality 7

$$-2p\sqrt{\lambda} < p^2 - \lambda(a^2-1) < 2p\sqrt{\lambda}, \tag{8}$$

where $p = \sqrt{1+\lambda(a+1)^2} - 1 > 0$. The inequalities 8 are equivalent to the following inequalities

$$p^2 + 2p\sqrt{\lambda} - \lambda(a^2-1) > 0, \tag{9}$$
$$p^2 - 2p\sqrt{\lambda} - \lambda(a^2-1) < 0. \tag{10}$$

We obtain from the inequalities 9 and 10, respectively:

$$p > (|a|-1)\sqrt{\lambda}, \tag{11}$$
$$(1-|a|)\sqrt{\lambda} < p < (1+|a|)\sqrt{\lambda}. \tag{12}$$

If $|a| > 1$ then we have from the inequalities 11-12

$$(|a|-1)\sqrt{\lambda} < p < (1+|a|)\sqrt{\lambda}. \tag{13}$$

We obtain from the inequality 13 eventually

$$(|a|-1)\sqrt{\lambda} + 1 < \sqrt{1+\lambda(a+1)^2} < 1 + (1+|a|)\sqrt{\lambda},$$
$$[(|a|-1)\sqrt{\lambda} + 1]^2 - 1 < \lambda(a+1)^2 < [1+(1+|a|)\sqrt{\lambda}]^2 - 1,$$
$$(|a|-1)^2\lambda + 2(|a|-1)\sqrt{\lambda} < \lambda(a+1)^2 < (1+|a|)^2\lambda + 2(1+|a|)\sqrt{\lambda},$$
$$-2|a|\lambda + 2(|a|-1)\sqrt{\lambda} < 2a\lambda < 2|a|\lambda + 2(1+|a|)\sqrt{\lambda},$$
$$-|a|\lambda + (|a|-1)\sqrt{\lambda} < a\lambda < |a|\lambda + (1+|a|)\sqrt{\lambda},$$
$$-|a| + (|a|-1)/\sqrt{\lambda} < a < |a| + (1+|a|)/\sqrt{\lambda},$$
$$(a-|a|)/(|a|+1) < 1/\sqrt{\lambda} < (a+|a|)/(|a|-1). \tag{14}$$

*The inequalities 14 are not valid for* $a \leq -e^{\alpha r_e}$.

If $a > 1$ then we have $1/\sqrt{\lambda} < 2a/(a-1)$ from the inequalities 14. Therefore *the inequalities 14 are not valid for* $a \geq 1/(1-2\sqrt{\lambda})$ *and* $\lambda < 1/4$.

If $0 < a < 1$ then we obtain eventually from the inequality 12

$$(1-a)\sqrt{\lambda} + 1 < \sqrt{1+\lambda(a+1)^2} < 1 + (1+a)\sqrt{\lambda},$$
$$[(1-a)\sqrt{\lambda} + 1]^2 - 1 < \lambda(a+1)^2 < [1+(1+a)\sqrt{\lambda}]^2 - 1,$$

$$(1-a)^2 \lambda + 2(1-a)\sqrt{\lambda} < \lambda(a+1)^2 < (1+a)^2 \lambda + 2(1+a)\sqrt{\lambda},$$

$$-2a\lambda + 2(1-a)\sqrt{\lambda} < 2a\lambda < 2a\lambda + 2(1+a)\sqrt{\lambda},$$

$$-a + (1-a)/\sqrt{\lambda} < a < a + (1+a)/\sqrt{\lambda},$$

$$1/\sqrt{\lambda} < 2a/(1-a). \tag{15}$$

It is easy to establish that *the inequality 15 is not valid for* $0 < a \leq 1/(1+2\sqrt{\lambda})$.

So we proved that *the Eq. 2 does not obey the inequality 5 in three cases:* 1) $a \leq -e^{\alpha r_e}$; 2) $a \geq 1/(1-2\sqrt{\lambda})$ *and* $\lambda < 1/4$; *and* 3) $0 < a \leq 1/(1+2\sqrt{\lambda})$. Therefore Eq. 2 and, hence, the solution of the Schrodinger equation, the analytical expressions for partition function, free energy, mean energy and specific heat, obtained in [1] from Eq. 2, are incorrect.

We finished the proof.

The following consideration confirms our above conclusions.

According to [1] Eq. 2 is valid for the vibrational quantum number $v$ which obeys the condition

$$0 \leq v \leq v_{max}, \tag{16}$$

$v_{max}$ is the most vibration quantum number, $v_{max} = v_+$ for $q < 0$ and $v_{max} = v_-$ for $q > 0$,

$$v_{\pm} = \left[ \left( \pm \sqrt{\lambda(a^2 - 1)} - \left(1 \pm \sqrt{1 + \lambda(a+1)^2}\right) \right)/2 \right]_{integer}, \tag{17}$$

where $[X]_{integer}$ denotes an integer part of $X$. But the condition 16 and Eq. 17 were not derived in [1], and therefore they need a justification.

It is easy to see that Eq. 17 is incorrect in the case a) $0 < a < 1$.

In the case when $q > 0$ Eq. 17 gives the inequality

$$v_{max} = \left[ -\left(\sqrt{\lambda(a^2-1)}\right) + 1 - \sqrt{1+\lambda(a+1)^2}\right)/2 \right]_{integer} < 0$$

is valid in the cases:

b) $a \leq -e^{\alpha r_e}$;
c) $a > (1+\lambda)/(1-\lambda)$ and $\lambda < 1$.

When $q < 0$ we have from Eq. 17 the inequality

$$v_{max} = \left[ \left(\sqrt{\lambda(a^2-1)} - 1 - \sqrt{1+\lambda(a+1)^2}\right)/2 \right]_{integer} < 0$$

is valid in the cases:

d) $a \geq 1$;
e) $-1 - 1/\lambda \leq a \leq -e^{\alpha r_e}$, $e^{\alpha r_e} < 1 + 1/\lambda$;
f) $a < -1 - 1/\lambda$, $e^{\alpha r_e} \leq 1 + 1/\lambda$, $\lambda \leq 1$;
g) $a \leq -e^{\alpha r_e}$, $e^{\alpha r_e} > 1 + 1/\lambda$, $\lambda \leq 1$;
h) $-(\lambda+1)/(\lambda-1) < a < -1 - 1/\lambda$, $e^{\alpha r_e} \leq 1 + 1/\lambda$, $\lambda > 1$;
i) $a \leq -e^{\alpha r_e}$, $e^{\alpha r_e} > 2$, $\lambda = 1$

j) $-(\lambda+1)/(\lambda-1) < a \leq -e^{\alpha r_e}$, $e^{\alpha r_e} < (\lambda+1)/(\lambda-1)$, $\lambda > 1$.

Therefore in the cases from a) to j) the solution of the Schrodinger equation, energy spectrum, given by Eq. 2, analytical expressions for the partition function, free and mean energies and specific heat capacity of diatomic molecule, obtained in [1] from Eq. 2, loss their physical sense.

The partition function $Q$ of the vibrational states of diatomic molecule was calculated in [1] from $Q = \sum_{v=0}^{v_{max}} e^{-E_{v,0}/kT}$, where $k$ is Boltzmann constant and $T$ is the temperature.

Eq. 2 gives $E_{v_{max}, J=0} = 0$ in the case when

$$\pm\sqrt{\lambda(a^2-1)} - \left(1 \pm \sqrt{1+\lambda(a+1)^2}\right) = 2n,$$

where $n = 0, 1, 2...$. The value $E_{v_{max}, J=0} = 0$ corresponds to the un-bonded state of the diatomic molecule. Therefore Eq. 17 cannot be used in order to calculate the partition function of bonded vibrational states of the diatomic molecule, and it is necessary replace Eq. 17 by the correct one. For example, Eq. 17 can be replaced by

$$v_{\pm} = \left[\pm\frac{1}{2}\sqrt{\lambda(a^2-1)} - \frac{1}{2}\left(1 \pm \sqrt{1+\lambda(a+1)^2}\right) - \frac{1}{2}\right]_{integer}$$

in order to avoid the inclusion of the un-bonded state to the partition function.

**Conclusion**

It is shown that the exact solution of the Schrodinger equation for the molecules consisting of two atoms interacting via improved Tietz potential, the analytical expressions for the energy of the rotational-vibrational levels, partition function, free energy, mean energy and specific heat obtained in [1] are incorrect. Therefore the conclusions of [1] are not reliable, and they could be incorrect.